\documentclass{revtex4}

\usepackage{amssymb}        
\usepackage{amsmath}        
\usepackage{graphicx}
\usepackage{color}        
\usepackage{float}

\begin{document}
\title{Aschenbach effect for spinning particles in Kerr spacetime}
\author{Jafar Khodagholizadeh}
\email{gholizadeh@ipm.ir}
\affiliation{Farhangian University, P.O. Box 11876-13311, Tehran, Iran.}
    
\author{Volker Perlick}
\email{perlick@zarm.uni-bremen.de}
\affiliation{ZARM, University of Bremen, 28359 Bremen, Germany}

\author{Ali Vahedi}
\email{vahedi@khu.ac.ir}
\affiliation{Department of Physics, Kharazmi University, 
Mofateh Ave, P.O. Box 15614, Tehran, Iran}


\maketitle

\section{Abstract}
The orbital velocity profile of
circular timelike geodesics in the equatorial plane of a Kerr black
hole has a non-monotonic radial behavior, provided that the spin 
parameter $a$ of the black hole is bigger than a certain critical value 
$a_c \approx 0.9953 \, M$. Here the orbital velocity is measured with 
respect to the Locally Non-Rotating Frame (LNRF), and the non-monotonic 
behavior, which is known as the Aschenbach effect, occurs only for 
co-rotating orbits. Using the Mathisson-Papapetrou-Dixon equations 
for a massive spinning particle, we investigate the Aschenbach
effect for test particles with spin. In addition to the black-hole spin, 
the absolute value of the particle’s spin and its orientation (parallel or
anti-parallel to the black-hole spin) also play an important role for the 
Aschenbach effect. We determine the critical value $a_c$ of the spin 
parameter of the Kerr black hole where the Aschenbach effect sets in
as a function of the spin of the probe. We consider not only black
holes ($a^2 \le M^2$) but also naked singularities ($a^2>M^2$). 
Whereas for spinless (geodesic) particles the orbital velocity is always
monotonically decreasing if the motion is counter-rotating, we find that for
spinning particles in counter-rotating motion with anti-parallel spin around 
a naked singularity the orbital velocity is increasing on a certain radius 
interval.

\vspace{0.2cm}

PACS: 04.70.Bw   85.30.Sf
\section{Introduction}\label{sec:intro}
As no signal can reach us from inside a black hole, the only way of observing
a stationary black hole is by detecting its influence on matter or on light 
rays that come close to it. In particular, we may observe electromagnetic
radiation emitted by matter that orbits a black hole in an accretion disk.
As a first approximation, it is reasonable to assume that the particles in an
accretion disk move on geodesics; this is true if they have no internal degrees 
of freedom, if they are not influenced by the interaction with neighboring 
particles or by external  (non-gravitational) fields and if their self-gravity 
is negligible. 

For particles moving on circular geodesics in the equatorial plane of the 
Kerr spacetime, Aschenbach \cite{Aschenbach2004,Aschenbach2006} 
made an interesting observation. He found that the orbital velocity might become an \emph{increasing} function of the radius coordinate on some
radius interval. This is in contrast to circular geodesic motion in the 
Schwarzschild metric, and also to circular motion in the Newtonian
$1/r$ potential, where the orbital velocity is always a \emph{decreasing} 
function of the radius coordinate, see e.g. Shapiro and 
Teukolsky  \cite{Shapiro}. More precisely, Aschenbach found that
this non-monotonic behavior of the orbital velocity occurs only if the spin
parameter, $a$, of the black hole satisfies an inequality $|a| \ge a_c$,
where the critical value $a_c \approx 0.9953 \, M$ is close to the
value of an extremal black hole, $|a|=M$, which characterizes the transition 
to a naked singularity. Moreover, the interval on which the orbital velocity is 
increasing occurs only for co-rotating, not for counter-rotating, orbits and 
it is close to but outside of the innermost stable circular orbit. Aschenbach 
related the non-monotonic behavior of the orbital velocity to the occurrence 
of certain resonances that could be observed (and, possibly, already have 
been observed with a few stellar black holes) as peaks in the power spectrum 
of the emitted radiation, see in particular Section 3 of \cite{Aschenbach2004}. 
If the interpretation is correct, the observation of those peaks gives direct 
information on the spin of the black hole.  

The non-monotonic behavior of the orbital velocity, called the 
\emph{Aschenbach effect} for short, has also been discussed for 
(non-geodesic) motion with constant specific angular momentum in
the Kerr spacetime \cite{StuchlikSlanyToeroekAbramowicz2005},
for geodesic motion in the Kerr-(anti-)de Sitter spacetime 
\cite{MuellerAschenbach2007,SlanyStuchlik2007} and in
braneworld generalizations of the Kerr spacetime \cite{zdenek2014},
and for the motion of charged particles in the field of a magnetized
Kerr black hole \cite{TursunovStuchlikKolos2016}.

Here we want to study the Aschenbach effect for particles with spin. To that end, 
we have to replace the geodesic equation with the Mathisson-Papapetrou-Dixon 
equations \cite{Mathisson1937, Papapetrou1951, Dixon1964}. As in the original 
work by Aschenbach, we restrict to motion in the equatorial plane of the Kerr 
spacetime. The spinning particle might be a rapidly rotating neutron star or a 
rapidly rotating hot spot in an accretion disk. In either case, the mass of the 
particle must be small enough to be negligible in comparison to the mass of 
the black hole. In the case of a neutron star orbiting a black hole, this puts, of 
course, limits on the applicability if the black hole has only a few Solar 
masses; for very massive stellar black holes, and for supermassive black 
holes, however, our analysis is applicable. It is our main goal to find out 
how the critical value of the black hole spin and the radius interval in which
the Aschenbach effect takes place is influenced by the particle's spin. 

The paper is organized as follows. In Section \ref{sec:MPD}
we recall some basic facts about the Mathisson-Papapetrou-Dixon 
equations. In Section \ref{sec:Kerr} we specialize to spinning particles 
in the equatorial plane of the Kerr metric, with the spin perpendicular
to this plane. Here it is our main goal to calculate the orbital velocity 
of circular orbits. To the best of our knowledge, this has not been 
done before, although there are numerous articles on such orbits, see
in particular the pioneering work by Rasband \cite{Rasband1973}
and by Tod et al. \cite{TodFeliceCalvani1976}. On the 
basis of the results from Section \ref{sec:Kerr}, we then discuss in 
Section \ref{sec:Aschenbach} the Aschenbach effect for spinning 
particles.

Our conventions are as follows. The signature of the metric is 
$(+,-,-,-)$ and we use units where the speed of light is $c=1$. 
We raise and lower indices with the spacetime metric, using 
Einstein's summation convention for greek indices running from 
0 to 3. Our index conventions for the curvature tensor are such
that  
\begin{equation}
R^{\tau}{}_{ \sigma \mu \nu}   =   \partial _{\mu} \Gamma ^{\tau} {}_{\nu \sigma}  -
\partial _{\nu} \Gamma ^{\tau} {}_{\mu \sigma}  +
\Gamma ^{\tau} {}_{\mu \rho} \Gamma ^{\rho} {}_{\nu \sigma}    -
\Gamma ^{\tau} {}_{\nu \rho} \Gamma ^{\rho} {}_{\mu \sigma}\, .
\label{eq:curv}
\end{equation}


\section{Mathisson-Papapetrou-Dixon equations}\label{sec:MPD}

On a spacetime with curvature tensor $R^{\mu}{}_{\nu \sigma \tau}$, 
the motion of a spinning extended body is determined in the pole-dipole 
approximation by the Mathisson-Papapetrou-Dixon equations 
\cite{Mathisson1937, Papapetrou1951, Dixon1964},
\begin{equation}
    \dfrac{D}{ds} x^{\mu} = u^{\mu} \, ,
\label{eq:MPD1}
\end{equation}
\begin{equation}\
    \dfrac{D}{ds} p^{\mu} = - \dfrac{1}{2} R^{\mu}{}_{\nu \sigma \tau}
    u^{\nu} S^{\sigma \tau} \, ,
\label{eq:MPD2}
\end{equation}
\begin{equation}
    \dfrac{D}{ds} S^{\mu \nu} = p^{\mu} u^{\nu}-p^{\nu} u^{\mu} \, .
\label{eq:MPD3}
\end{equation}
Here $x^{\mu} (s)$ is the worldline of a reference point inside the body,  
$u ^{\mu} (s)$ is the corresponding 4-velocity, and $\dfrac{D}{ds}$ denotes 
covariant derivative, in the direction of $u^{\mu} (s)$, of tensor fields along 
the worldline $x^{\mu} (s)$.  In a given spacetime background, this is a system 
of first-oder ordinary differential equations for the worldline $x^{\mu} (s)$, the 
momentum $p_{\nu}(s)$ and the spin tensor $S^{\mu \nu} (s) = 
- S^{\nu \mu} (s)$ of the particle. Note that this system of equations is 
invariant under arbitrary reparametrizations,
\begin{equation}
\dfrac{D}{ds} \mapsto k \, \dfrac{D}{ds} \, , \quad
u^{\mu} \mapsto k \, u^{\mu} \, , \quad
p_{\mu} \mapsto  p_{\mu} \, , \quad
S^{\mu \nu} \mapsto  S^{\mu \nu} 
\label{eq:reparam}
\end{equation}
where $k$ is a nowhere vanishing function of $s$. For our purpose, we find 
it convenient to choose the proper time parametrization,
\begin{equation}
g_{\mu \nu} u^{\mu} u^{\nu} = 1 \, .
\label{eq:proper}
\end{equation}
If $u^{\mu}$ and $p^{\mu}$ are timelike and future-oriented, which is usually 
required for physically reasonable solutions, we may define two real and positive 
quantities 
\begin{equation}
\mu := \sqrt{g_{\rho \sigma} p^{\rho} p^{\sigma}} \, , \quad
m := g_{\rho \sigma} u^{\rho} p^{\sigma} \, .
\label{eq:mmu}
\end{equation}
$\mu$ is the mass of the particle in the center-of-momentum system whereas $m$
is the mass in the rest system of an observer comoving along the worldline $x^{\rho} (s)$.
In general, neither $m$ nor $\mu$ is guaranteed to be a constant of motion.

As the system of Mathisson-Papapetrou-Dixon equations is underdetermined, we have
to add a \emph{supplementary condition}
\begin{equation}
    V^{\rho} S_{\rho \sigma} = 0 \, .
\label{eq:Suppl}
\end{equation}
Here $V^{\rho}$ is a timelike vector field along the worldline $x^{\mu}(s)$
we are free to choose at will. For convenience, we will assume $V^{\rho}$ to 
be normalized according to $V_{\rho}V^{\rho}=1$. The most common choices 
for $V^{\rho}$ are $V^{\rho}= p^{\rho}/\mu$ (Tulczyjew-Dixon 
condition \cite{Tulczyjew1959, Dixon1970}) and $V^{\rho}=u^{\rho}$ 
(Frenkel-Mathisson-Pirani condition \cite{Frenkel1926,Mathisson1937, Pirani1956}). 
It is well known that $\mu$ is a constant of motion if the Tulczyjew-Dixon 
condition is imposed whereas $m$ is a constant of motion if the 
Frenkel-Mathisson-Pirani condition is imposed. 

As soon as we have fixed the vector field $V^{\mu}$, we can express the spin
tensor $S_{\mu \nu}$ in terms of a spin vector $S^{\rho}$,
\begin{equation}
    S_{\mu \nu} = \varepsilon _{\mu \nu \sigma \rho} V^{\sigma} S^{\rho} 
    \, , \quad
    S_{\rho} V^{\rho} = 0 \, ,
\label{eq:spinvec}
\end{equation}
where $\varepsilon _{\mu \nu \sigma \rho}$ is the totally antisymmetric Levi-Civita
tensor field (volume form) of the spacetime metric.   
 
The ambiguity in choosing a supplementary condition is understood if we recall
that a body with a given spin different from zero must have a minimum size, just to make sure that no parts of the body move at a superluminal speed \cite{Moller1949}.  
Choosing a supplementary condition corresponds to choosing a particular worldline 
$x^{\mu} (s)$ inside the worldtube of such a finite-size body.   
 
\section{Spinning particle in the equatorial plane of the Kerr spacetime}\label{sec:Kerr}

We now specify the background metric to the Kerr metric which reads, in standard 
Boyer-Lindquist coordinates, \cite{BoyerLindquist1967}
\begin{equation}
g_{\mu \nu} dx^{\mu} dx^{\nu}
=  
\left(1-\frac{2Mr}{\rho^2}\right) dt^2 
- 
\frac{\rho^2}{\Delta} dr^2 
- 
\rho^2 d\vartheta^2 
-
\mathrm{sin}^2\vartheta\left(r^2+a^2+
\frac{2Mra^2\mathrm{sin}^2\vartheta}{\rho^2}\right)
d\varphi^2
+ 
\frac{4Mra \mathrm{sin} ^2\vartheta}{\rho^2} \, dt \, d\varphi
\label{eq:kerr}
\end{equation}
where
\begin{equation}
\rho  ^2:= r^2+a^2 \mathrm{cos}^2\vartheta
\, , \quad
\Delta  = r^2+a^2-2Mr \, .
\label{eq:rhoDelta}
\end{equation}
Here $M$ is the mass parameter and $a$ is the spin parameter. Both
have the dimension of a length. For $a^2 \leq M^2$ we have a black
hole whereas for $a^2 > M^2$ we have a naked singularity.

We want to consider the Mathisson-Papapetrou-Dixon equations with a 
supplementary condition (\ref{eq:Suppl}), where for the time being 
$V^{\mu}$ is specified only to be the 4-velocity field of observers in
circular motion,
\begin{equation}
V^{\mu} \partial _{\mu} = V^t \partial _t + V^{\varphi} \partial _{\varphi}
\, .
\label{eq:supplcirc}
\end{equation}
We are interested in circular motion in the equatorial plane,
\begin{equation}
\vartheta = \pi /2 \, , \quad
u^{\mu} \partial _{\mu} = u^t \partial _t + u^{\varphi} \partial _{\varphi}
\, ,
\label{eq:eqcirc}
\end{equation}
with the spin perpendicular to the equatorial plane,
\begin{equation}
S^{\mu} \partial _{\mu} = S^{\vartheta} \partial _{\vartheta} 
\, , \quad
S^{\vartheta} = - \, \dfrac{S}{r} \, .
\label{eq:spinperp}
\end{equation}
Here $S$ is a constant of motion,
\begin{equation}
g_{\mu \nu} S^{\mu} S^{\nu} = - S^2 \, ,
\label{eq:S}
\end{equation}
that may be positive or negative. We have $aS>0$ if the spin of the particle is
parallel to the spin of the black hole and $aS<0$ if it is anti-parallel.  

Under these assumptions, evaluating all components of the Mathisson-Papapetrou-Dixon 
equation (\ref{eq:MPD3}) yields
\[
p^r =0 \, , \quad p^{\vartheta} = 0 \, , 
\]
\begin{equation}
- S u ^{\varphi} V ^{\varphi}   + \dfrac{MS}{r^3} \big( u^t-a u^{\varphi} \big) 
\big( V^t - a V^{\varphi} \big) = 
\big( p^t - a p^{\varphi} \big) u^{\varphi}  - p ^{\varphi} \big(u^t-a u^{\varphi} \big)
\, .
\label{eq:Eq1}
\end{equation}
Similarly, from  (\ref{eq:MPD2}) we find
\[
\dfrac{dp^t}{ds}= 0  \, , \quad
\dfrac{dp^{\varphi}}{ds}= 0  \, , 
\]
\[
M \, r^2 \big( p^t - a p^{\varphi} \big)  \big( u^t - a u^{\varphi} \big) 
-r^5  p^{\varphi}  u^{\varphi} 
\]
\begin{equation}
=    -3 MS a \big( u^t- a u^{\varphi} \big) \big( V^t-a V^{\varphi} \big) + 
MS r^2 \Big( 2 ( u^t- a u^{\varphi} ) V^{\varphi} + u^{\varphi} (V^t-a V^{\varphi} ) \Big) 
\, .
\label{eq:Eq2}
\end{equation}

\subsection{Tulczyjew-Dixon condition}
If the Tulczyjew-Dixon supplementary condition $V^{\rho} = p^{\rho} / \mu$
is imposed, $\mu$ is a constant of motion and it is convenient to characterize the
particle's spin by the dimensionless parameter
\begin{equation}
s = \dfrac{S}{M \, \mu} \, .
\label{eq:defs}
\end{equation}
Note that, according to the notation of (\ref{eq:MPD1}), (\ref{eq:MPD2}) and (\ref{eq:MPD3}),
our solutions to the Mathisson-Papapetrou-Dixon equations are parametrized by a curve 
parameter which was also denoted $s$. The latter, however, will not explicitly appear any
more, so there is no danger of confusion. 

Eqs. (\ref{eq:Eq1}) and (\ref{eq:Eq2}) specify to 
\begin{equation}
\dfrac{p^t - a p^{\varphi}}{p^{\varphi}} =
\dfrac{
u^t- a u^{\varphi}  -  s M u^{\varphi} 
}{
u^{\varphi} - \dfrac{M^2 s}{r^3} \big( u^t - a u ^{\varphi}\big)
}
\, ,
\label{eq:Eq1TD}
\end{equation} 
\begin{equation}
    \dfrac{p^t - a p^{\varphi}}{p^{\varphi}}   = 
\dfrac{
\dfrac{r^5 }{M} u^{\varphi} + 2 s M  r^2 ( u^t - a u^{\varphi} ) 
}{
r^2   ( u^t-a u^{\varphi} )   + 3 a s M ( u ^t - a u^{\varphi} )  
-  s M r^2  u ^{\varphi} 
}
\, . 
\label{eq:Eq2TD}
\end{equation} 
If we introduce the angular velocity
\begin{equation} 
\Omega = \dfrac{u^{\varphi}}{u^t} 
\label{eq:Omega}
\end{equation}
equating the right-hand sides of (\ref{eq:Eq1TD}) and (\ref{eq:Eq2TD}) yields 
\begin{equation} 
\left( 1 + \dfrac{3as M}{r^2 } + \dfrac{2M^3 s^2}{r^3} \right) 
\big(\Omega ^{-1} - a \big )^2
- 3 s M  \left( 1 + \dfrac{as M}{r^2}\right) \big(\Omega ^{-1} - a \big )
- \dfrac{r^3}{M} + s^2 M^2 = 0 \, .
\label{eq:quadTD}
\end{equation}
This is a quadratic equation for $(\Omega ^{-1} -a )$ with solutions 
\begin{equation} 
\Omega _{\pm} ^{-1} - a 
= 
\dfrac{
3 M^2 r^3 s  + 3 a M^3  r s^2 \pm \sqrt{M} \, r \, 
\sqrt{D}
}{
2 M r^3 + 6 a M^2 r  s + 4 M^4  s^2
}
\label{eq:OmegaTD}
\end{equation}
with
\begin{equation}
D = 4r^7+12Mar^5s+13M^3r^4s^2+6M^4ar^2s^3+(9a^2-8Mr) M^5 s^4
\, .
\label{eq:defD}
\end{equation}
Note that, because of the normalization condition (\ref{eq:proper}),
we have
\begin{equation}
g_{tt} \Omega ^{-2} + 2 g_{t \varphi} \Omega ^{-1}
+ g_{\varphi \varphi} = \dfrac{1}{\big( u^{\varphi} \big)^2} 
\, .
\label{eq:uphi} 
\end{equation}
After inserting the metric coefficients the condition of 
$1/ \big( u ^{\varphi} \big) ^2 > 0$ requires that 
\begin{equation}
\Big( 1- \dfrac{2M}{r} \Big) \big( \Omega ^{-1} - a \big) ^2 
+ 2 a \big( \Omega ^{-1} - a \big) 
-  r^2   > 0 \, .
\label{eq:sublum}
\end{equation}
This inequality makes sure that the 4-velocity of the particle is 
timelike, i.e., that the motion is subluminal. If, at a certain radius 
value $r$, the discrimant $D$ defined in 
(\ref{eq:defD}) is negative, then there is no solution to our motion
problem at this radius value. If $D$ is non-negative, there may be 
two solutions (typically one co-rotating and the other counter-rotating), 
one solution or no solution, depending on whether (\ref{eq:sublum}) 
is satisfied for both $\Omega = \Omega _+$ and $\Omega = 
\Omega _-$, only for one of them, or for neither of them.
 
The angular velocity $\Omega$ is a useful auxiliary quantity
from a mathematical point of view, but it is not a physically
meaningful quantity, at least not in the region we are interested
in. It describes the motion with respect to the vector field
$\partial _t$ which is not timelike inside the ergoregion, i.e.,
in the domain which is of relevance to the Aschenbach effect. 
Therefore, $\Omega$ is not
the angular velocity with respect to an observer field. For describing
the motion with respect to an observer field, we introduce the 
orbital velocity with respect to the Locally Non-Rotating 
Frame (LNRF) \cite{zamo,LNRF}
\begin{equation}
e_0=
\dfrac{
-g_{\varphi \varphi} \partial _t + g_{t \varphi} \partial {\varphi}
}{
\sqrt{-g_{\varphi \varphi} \big( g_{t \varphi} ^2 - g_{\varphi \varphi} g_{tt}  \big)}
} 
\, , \quad
e_1 =
\dfrac{\partial _r}{\sqrt{-g_{rr}}}
\, , \quad
e_2 =
\dfrac{\partial _{\vartheta}}{\sqrt{-g_{\vartheta \vartheta}}}
\, , \quad
e_3 =
\dfrac{\partial _{\varphi}}{\sqrt{-g_{\varphi \varphi}}}
\, .
\label{eq:LNRF}
\end{equation}
This is an orthonormal tetrad if $\Delta > 0$, i.e., everywhere except between the 
two horizons. Observers with 4-velocity $e _0$ are also known as Zero Angular
Momentum Observers (ZAMOs). 

For circular motion, the orbital velocity $\mathcal{V}$ with respect to the 
LNRF is determined by
\begin{equation}\label{eq:Vdef}
u^t \partial _t + u^{\varphi} \partial _{\varphi} =
N \Big( e_0 + \mathcal{V} \, e_3 \Big) 
\end{equation}
where $N$ is a scalar factor. For timelike orbits $\mathcal{V}$ takes values
between $-1$ and 1. Comparing coefficients of $\partial _t$ and 
$\partial _{\varphi}$ in (\ref{eq:Vdef}) allows us to express 
$\Omega = u^{\varphi}/u^t$ in terms of $\mathcal{V}$,
\begin{equation}
\Omega = 
\dfrac{
g_{t \varphi} + \mathcal{V} \, 
\sqrt{g_{t \varphi} ^2 - g_{\varphi \varphi} g_{tt}}
}{-g _{\varphi \varphi}}
\, .
\label{eq:OmegaV}
\end{equation}
After inserting the metric coefficients and solving for $\mathcal{V}$ we find
\begin{equation}
\mathcal{V} 
=
 \dfrac{\big((r^2+a^2)^2-a^2 \Delta \big)\Omega-2aMr}{r^2 \, \sqrt{\Delta}}  
\, .
\label{eq:VOmega}
\end{equation}
With $\Omega=\Omega _{\pm}$ from (\ref {eq:OmegaTD}), this gives us 
two solutions for the orbital velocity,
\begin{equation}
\mathcal{V{}_{\pm}} 
= 
- 
\dfrac{2Ma}{r \sqrt{\Delta}} 
+
\dfrac{
\Big( 2Mar^3+3M^2r(2a^2+r^2)s+M^3a(4M+3r)s^2 \mp \sqrt{M} r 
\sqrt{D}
\Big)
\big(2Ma^2+a^2r+r^3 \big)
}{
\Big( 2Ma^2r^3-2r^6+6M^2ar(a^2+r^2)s+2M^3(r^3+3a^2r+2Ma^2)s^2 \Big)
r \sqrt{\Delta}
}
\, .
\label{eq:VTD}
\end{equation}
For the existence of an orbit with velocity $\mathcal{V}_+$ (or $\mathcal{V}_-$,
respectively) at radius value $r$ it is necessary and sufficient that $D$ is
non-negative and that $\big| \mathcal{V}{}_+ \big| < 1$ (or 
$\big| \mathcal{V}{}_- \big| < 1$, respectively).

Far away from the center, (\ref{eq:OmegaTD}) and (\ref{eq:VTD}) may
be approximated as
\begin{equation}
\Omega{}_{\pm}
= 
\, \pm \, \dfrac{\sqrt{M}}{\sqrt{r^3}} 
\Big( 1 + O \big( (M/r)^{1/2} \Big)
\, ,
\label{eq:asyOmega}
\end{equation}
\begin{equation}
\mathcal{V}{}_{\pm}
= 
\pm \dfrac{\sqrt{M}}{\sqrt{r}} \Big( 1 + O \big( (M/r)^{1/2} \Big)
\, .
\label{eq:asyV}
\end{equation}
From these equations, we read that, for any choice of $a$ and $S$,
there are two circular orbits at all sufficiently large radius values;
by (\ref{eq:asyV}), the label + refers to an orbit with positive $\mathcal{V}$, 
i.e., a particle moving in the positive $\varphi$ direction with respect to 
the ZAMOs, whereas the label - refers to an orbit with negative 
$\mathcal{V}$, i.e., a particle moving in the negative $\varphi$ direction
with respect to the ZAMOs. This means that for $a>0$ the + orbit
is co-rotating and the - orbit is counter-rotating; for $a<0$ it is 
vice versa. Far away from the center, $\mathcal{V}{}_{+}$ goes
monotonically to zero from above and $\mathcal{V}{}_{-}$ goes
monotonically to zero from below. It is the subject of this paper to 
investigate if and how this monotonic behavior changes closer to
the central object.  We will see that $\mathcal{V}{}_+$ and 
$\mathcal{V}{}_-$ may change sign so that it is not always 
true that for $a>0$ the + orbit is co-rotating and the - orbit
is counter-rotating. Also, $\big| \mathcal{V}{}_+ \big|$ and 
$\big| \mathcal{V}{}_- \big|$ may become bigger than 1; in
regions where this happens the corresponding orbit does not
exist at all.
   
Taylor expansion with respect to the spin parameter $s$ of 
(\ref{eq:OmegaTD}) and (\ref{eq:VTD}) yields
\[
\Omega{}_{\pm}   
= 
\dfrac{\sqrt{M}}{a \sqrt{M} \pm \sqrt{r^3}}
\pm
\dfrac{3 \sqrt{M^3} \big( a  \mp \sqrt{Mr} \big) 
}{
2 \sqrt{r} \Big( a \sqrt{M} \pm \sqrt{r^3} \Big)^2
} \, s
\]
\begin{equation}
\pm
\dfrac{3 M^2 
\Big(
(M-4r) a \sqrt{M} r  - 9a^3 \sqrt{M} 
\pm (8M-3r) a^2 \sqrt{r} \pm 7 M \sqrt{r^5} 
\Big) }{
8 r^2 \Big( a \sqrt{M} \pm \sqrt{r^3} \Big) ^3 }
\, s^2 + O \big( s^3 \big) 
\label{eq:OmegaTaylor}
\end{equation}
and
\[
\mathcal{V}{}_{\pm}
=
\dfrac{
\sqrt{M} \Big( r^2+a^2  \mp 2a \sqrt{ M r}\Big)
}{
\sqrt{\Delta} \Big( a \sqrt{M} \pm \sqrt{r^3} \Big)
}
\pm
\dfrac{
3  \sqrt{M^3} \Big( r^3+a^2 r+2Ma^2 \Big) \Big( a \mp  \sqrt{Mr} \Big)
}{
2 \sqrt{r^3} \sqrt{\Delta} \Big( a \sqrt{M} \pm \sqrt{r^3} \Big)^2
}
\, s 
\]
\begin{equation}
\pm
\dfrac{
3 \, \sqrt{M^5}
\Big(
(M-4r) a \sqrt{M} r  - 9a^3 \sqrt{M} \pm (8M-3r) a^2 \sqrt{r} \pm 7 M \sqrt{r^5} 
\Big)
 \Big(r^3+2Ma^2+a^2r \Big)
 }{
8 \, \sqrt{r^7} \,  \sqrt{\Delta} \big( a \sqrt{M} \pm \sqrt{r^3} \big)^3
}
\, s^2
+ O \big( s^3 \big) \, . 
\label{eq:VTaylor}
\end{equation}
For vanishing spin, $s=0$, we recover the well-known equations
for circular geodesics.

When we discuss the Aschenbach effect in Sec. \ref{sec:Aschenbach} 
we have to make sure that the circular orbits in question are stable
because otherwise they would hardly be realized in Nature.
For this reason we need to know the position of the Innermost
Stable Circular Orbit (ISCO). The latter can be calculated with the 
help of an effective potential which can also be used for checking 
whether our equation (\ref{eq:quadTD})  is in agreement with
the results of other authors on the subject. It is well known
that the radial component of the Mathisson-Papapetrou-Dixon
equations with the Tulczyjew-Dixon supplementary condition
can be characterized by an effective potential $U_{E,J_z}(r)$,
say, which depends on the constants of motion $E$ and $J_z$
associated with the Killing vector fields $\partial _t$ and
$\partial _{\varphi}$, respectively. For the precise form 
of this function $U_{E,J_z}(r)$ we refer to Saijo et 
al. \cite{SaijoEtAl1998} where it is given by the first
three terms on the left-hand
side of eq. (2.26). The circular orbits are determined by
solving simultaneously the equations  $U_{E,J_z}(r)=0$ and
$dU_{E,J_z}(r)/dr =0$. As $E$ and $J_z$ are in a one-to-one
correspondence with the velocity components $u^t$ and
$u^{\varphi}$ (see eq. (2.10) in Ref. \cite{SaijoEtAl1998}), these two
equations determine $u^t$ and $\Omega = u^{\varphi}/u^t$
as functions of $r$. It is straight forward to verify that,
after eliminating $u^t$, this results in our equation  (\ref{eq:quadTD})
for $\Omega$, so we see that the latter is indeed in agreement
with the known characterization of the circular orbits in
terms of the effective potential.  For determining the
radius coordinate of the ISCO we have to solve the three
equations $U_{E,J_z}(r)=0$, $dU_{E,J_z}(r)/dr =0$ and
$d^2U_{E,J_z}(r)/dr^2 =0$ for $r$, $E$ and $J_z$. This has 
already been done numerically by various authors. If the radius 
coordinate of the ISCO has
been found, its angular velocity can be determined with the
help of our eq.  (\ref{eq:quadTD}). We have done this numerically 
for several values of $a/M$ and $s$ and compared with Table I
and Table II in Lukes-Gerakopoulos et al. \cite{LukesEtAl2017}. 
We have found agreement up to five digits, thereby confirming 
that our equation (\ref{eq:quadTD}) is in accordance with 
previous results of others.

\subsection{Frenkel-Mathisson-Pirani condition}
If the Frenkel-Mathisson-Pirani supplementary condition $V^{\rho} = u^{\rho}$
is imposed, Eq. (\ref{eq:Eq1}) implies
\begin{equation}
p^t-ap^{\varphi} = m \big( u^t-a u^{\varphi} \big) + r^2 S \big( u^{\varphi} \big)^3
-aS \big( u^{\varphi} \big)^2 \big(u^t-a u^{\varphi} \big)
- \dfrac{M}{r} S u^{\varphi} \big( u^t - a u ^{\varphi} \big) ^2
+ \dfrac{aM}{r^3} S \big( u^t-a u^{\varphi} \big) ^3
\label{eq:puFMP1}
\end{equation}
and
\begin{equation}
p^{\varphi} = m \,  u^{\varphi}  + 
+ a S ( u^{\varphi} ) ^3
+ \Big( 1- \dfrac{2M}{r} \Big) S ( u ^{\varphi} ) ^2 ( u^t - a u ^{\varphi} ) 
-\dfrac{aMS}{r^3} u^{\varphi}  ( u^t - a u ^{\varphi} ) ^2
- \Big( 1- \dfrac{2M}{r} \Big) \dfrac{MS}{r^3} ( u^t - a u ^{\varphi} ) ^3
\, .
\label{eq:puFMP2}
\end{equation}
In this case $m$ is a constant of motion, so we characterize the particle's spin
by the dimensionless parameter
\begin{equation}
\tilde{s} = \dfrac{S}{Mm} \, .
\end{equation}
Then inserting (\ref{eq:puFMP2}) into (\ref{eq:Eq2}) and using (\ref{eq:uphi}) yields
a fourth-order equation for  $\Omega ^{-1} -a$,
\[
\left( 1 - \dfrac{2M}{r} + \Big( \dfrac{3a}{r^2}-\dfrac{5aM}{r^3} \Big) M \tilde{s}  \right)
\big( \Omega ^{-1} -a \big) ^4
+ \left( 2 a + \Big(\dfrac{3M}{r}-2+\dfrac{6a^2}{r^2} \Big) M \tilde{s} \right)
\big( \Omega ^{-1} -a \big) ^3
\]
\begin{equation}
+ \left( r^2 - \dfrac{r^3}{M}-9a M \tilde{s}  \right) 
\big( \Omega ^{-1} -a \big) ^2
- \left( \dfrac{2ar^3}{M} - \Big( 6 M r^2- r^3 \Big) \tilde{s}  \right)
\big( \Omega ^{-1} -a \big)
+ \dfrac{r^5}{M} - ar^3 \tilde{s}  =0 
\, .
\label{eq:quartFMP}
\end{equation} 
This equation can be analytically solved for $\Omega ^{-1} -a$, using a standard
method for solving a fourth-order equation, but the resulting expressions are quite 
awkward and will not be given here. If we Taylor expand with respect to $\tilde{s}$, we find 
that the four solutions $(\Omega _+, \Omega _-, \hat{\Omega} _+,\hat{\Omega} _-)$
are  
\begin{equation}
\Omega{}_{\pm}   
= 
\dfrac{\sqrt{M}}{a \sqrt{M} \pm \sqrt{r^3}}
\pm
\dfrac{3 \sqrt{M^3} \big( a  \mp \sqrt{Mr} \big) 
}{
2 \sqrt{r} \, \big( a \sqrt{M} \pm \sqrt{r^3} \big)^2
}
\, \tilde{s} 
+
\dfrac{3 \sqrt{M^5}
\Big(
6 \, r^2  \big( a \mp \sqrt{Mr} \big) ^2
-
\big( a \sqrt{M} \pm \sqrt{r^3} \big) \, K_{\pm}
\Big)
}{
8 \, r^3 \, \big(a \sqrt{M} \pm \sqrt{r^3} \big)^3
}
\, \tilde{s}{}^2 + O \big( \tilde{s}{}^3 \big) 
\label{eq:OmegaFMP}
\end{equation}
where 
\begin{equation}
K_{\pm}
=
\dfrac{
(9r-43M) a^2 r+(3M-r) M r^2 \pm 2 (9a^2+11 Mr-4 r^2 ) a \sqrt{Mr}
}{
\Big( 2 a  \sqrt{M} \pm \sqrt{r} \big( r- 3 M \big)  \Big) 
}
\end{equation}
and
\begin{equation}
\hat{\Omega}{}_{\pm}   
=
\dfrac{2M-r}{2aM\pm r \sqrt{\Delta}}
+
\dfrac{ 
M \Big( (r^2-5Mr+6M^2 ) r-2 a M \big( a \pm \sqrt{\Delta} \big)  \Big) 
}{
2r  (2 a M \pm r \sqrt{\Delta} )^2
}
\, \tilde{s} 
+
\dfrac{ M^3 \Big( \sqrt{\Delta} \, Q \pm 2 M P \Big)}{
8 M r^3 \sqrt{\Delta} \Big( 2 a M \pm r \sqrt{\Delta} \Big) ^3
}
\, \tilde{s}{}^2 + O \big( s^3 \big)
\, ,
\label{eq:hatOmegaFMP}
\end{equation} 
where
\begin{equation}
Q =
8 a^4 M (5 M+3 r)-8 a^2 M r \big( 9 M^2+6 M r - 5 r^2 \big)
+ r^3 \big( 66 M^3-69 M^2 r +20 M r^2 - r^3 \big)
\end{equation}
and
\begin{equation}
P =
4 a^5 \big( 5M + 3 r \big)-2 a^3 r \big( 28 M^2+13 M r-13 r^2 \big)
+3 a r^2 \big( 10 M^3+11 M^2 r-18 M r^2+5 r^3 \big)
\, .
\end{equation}
The corresponding orbital velocities read
 \[
\mathcal{V}{}_{\pm}
=
\dfrac{
\sqrt{M} \Big( r^2+a^2 \mp 2a \sqrt{ M r}\Big)
}{
\sqrt{\Delta} \Big( a \sqrt{M} \pm \sqrt{r^3} \Big)
}
\pm
\dfrac{
3  \sqrt{M^3} \Big( r^3+a^2 r+2Ma^2 \Big) \Big( a \mp  \sqrt{\Delta} \Big)
}{
2 \sqrt{r^3} \sqrt{\Delta} \Big( a \sqrt{M} \pm \sqrt{r^3} \Big)^2
}
\, \tilde{s} 
\]
\begin{equation}
+
\dfrac{
3 \sqrt{M^5} (r^3+a^2 r+2 M a^2 ) \Big( 
3 (a \mp \sqrt{M r} )^2
- (a \sqrt{M} \pm \sqrt{r^3} ) K_{\pm}
\Big)
}{
4 r^2 (a \sqrt{M} \pm \sqrt{r^3} ) ^3  \sqrt{\Delta}
}
\, \tilde{s}{}^2
+ O \big( \tilde{s}{}^3 \big)
\label{eq:VFMP}
\end{equation}
and
\[
\hat{\mathcal{V}}{}_{\pm}
=
\, \mp \, 1
+
\dfrac{
M \big( r^3+a^2 r+2Ma^2 \big) \Big( r^3-5Mr^2+6M^2r-2aM \big( a \pm  \sqrt{\Delta} \big) \Big)
}{
2 r^2  \sqrt{\Delta} \Big( 2 M a  \pm r  \sqrt{\Delta} \Big) ^2
}
\, \tilde{s} 
\]
\begin{equation}
\pm
\dfrac{
M^2  \big( r^3+a^2 r+2 M a^2 \big) \big( 2 M P \pm \sqrt{\Delta} \, Q \big) 
}{
8 r^4 \Delta  (2 a M \pm r \sqrt{\Delta} )^3
}
\, \tilde{s}{}^2
+ O \big( \tilde{s}{}^3 \big) \, .
\end{equation}
We see that the hatted solutions are unphysical because 
$ \big| \mathcal{V}{}_{\pm} \big|$ becomes the velocity of light for 
$\tilde{s} \to 0$, both for the $+$ and the $-$ branch. The orbital 
velocity is even superluminal for small spin values of one sign. This 
observation was made already by Costa et al. \cite{CostaLukesSemerak2018} 
where the four exact solutions for the angular velocity are 
worked out, in the supplemental material, for the Schwarzschild 
solution. Their result is in agreement with our eq. (\ref{eq:quartFMP})
if the latter is specified to the Schwarzschild case $a=0$. If we discard the 
hatted solutions, we are left with two solutions, given in (\ref{eq:OmegaFMP}) 
and (\ref{eq:VFMP}), that have, indeed, the correct geodesic limit for 
$\tilde{s} \to 0$. They coincide not only to zeroth but also to first order 
with the solutions from the Tulczyjew-Dixon condition, see 
(\ref{eq:OmegaTaylor}) and (\ref{eq:VTaylor}), where $s$ in the 
Tulczyjew-Dixon case has to be replaced by $\tilde{s}$ in the 
Frenkel-Mathisson-Pirani case. The second and higher-order 
terms, however, are different. This is in agreement with a more 
general result that can be read from Section 2 of Chicone et 
al. \cite{ChiconeMashhoonPunsly2005}: To within linear approximation 
with respect to the spin, on any spacetime the Tulczyjew-Dixon condition 
is equivalent to the Frenkel-Mathisson-Pirani condition, and the spin 
parameters $s$ and $\tilde{s}$ (in our notation) actually coincide.

At the end of this section we will again indicate how to
check stability of the circular orbits, i.e., how to calculate
the ISCO. Whereas this could be done with the help of
\emph{one} effective potential for the Tulzcyjew-Dixon
condition, we need \emph{three} potentials in the case
of the Frenkel-Mathisson-Pirani condition. For characterizing
the circular orbits one has to equate the potentials and their
first $r$-derivatives to zero, see eqs. (45) in Harms et al.
\cite{HarmsEtAl2016}. These are six equations
for six unknowns which include $u^t$ and $u^{\varphi}$
(in our notation) or, equivalently, $u^t$ and $\Omega =
u^{\varphi}/u^t$. It is straight-forward, though somewhat
tedious, to arrive at an equation for $\Omega$ by eliminating
the other five unknowns. This procedure results, indeed, in
our equation (39) which demonstrates that this equation is in
agreement with the characterization of the circular orbits in
terms of the three effective potentials. For calculating the
ISCO we have to equate the three potentials together with
their first and second derivatives to zero, as outlined by
Harms et al., which results in nine equations for nine
unknowns. From these equations one can numerically determine
the radius coordinate of the ISCO. We have done this, for
several values of $a/M$ and $\tilde{s}$, and then calculated
the angular velocity of the ISCO with our eq. (39), choosing
the unhatted solutions. Again, we have found agreement up to
five digits with the values given in Table I and Table II
of Lukes-Gerakopoulos et al. \cite{LukesEtAl2017}
.
\section{Aschenbach effect in the equatorial plane of 
         Kerr spacetime}\label{sec:Aschenbach}

We consider orbits of spinning particles with the Tulczyjew-Dixon 
condition, i.e. with orbital velocity given by (\ref{eq:VTD}). To 
within linear approximation with respect to the spin, the results 
are valid also for the Frenkel-Mathisson-Pirani condition. Without loss of 
generality, we assume $a>0$. Then $s= S/(M \mu)$ is positive if 
the particle's spin is parallel to the spin of the black hole and 
negative if it is anti-parallel.

Our first task is to find out for which values of the relevant 
parameters the Aschenbach effect occurs, i.e., for which values 
of $a$ and $s$ there is an interval of the radius coordinate on 
which the orbital velocity is increasing with the radius coordinate. 
We first consider the black-hole case, $a \le M$, and we restrict 
to the domain of outer communication, i.e. to the region outside of 
the outer horizon. By inspection, we find that $| \mathcal{V}{}_- |$ 
is always monotonically decreasing with $r$ on this domain, so we 
only have to discuss $\mathcal{V}{}_+$. On the considered
domain, $\mathcal{V}{}_+$ is always positive, i.e., it describes  
co-rotating orbits. 

\begin{figure}[H]
\begin{center} 
\includegraphics[width=11cm]{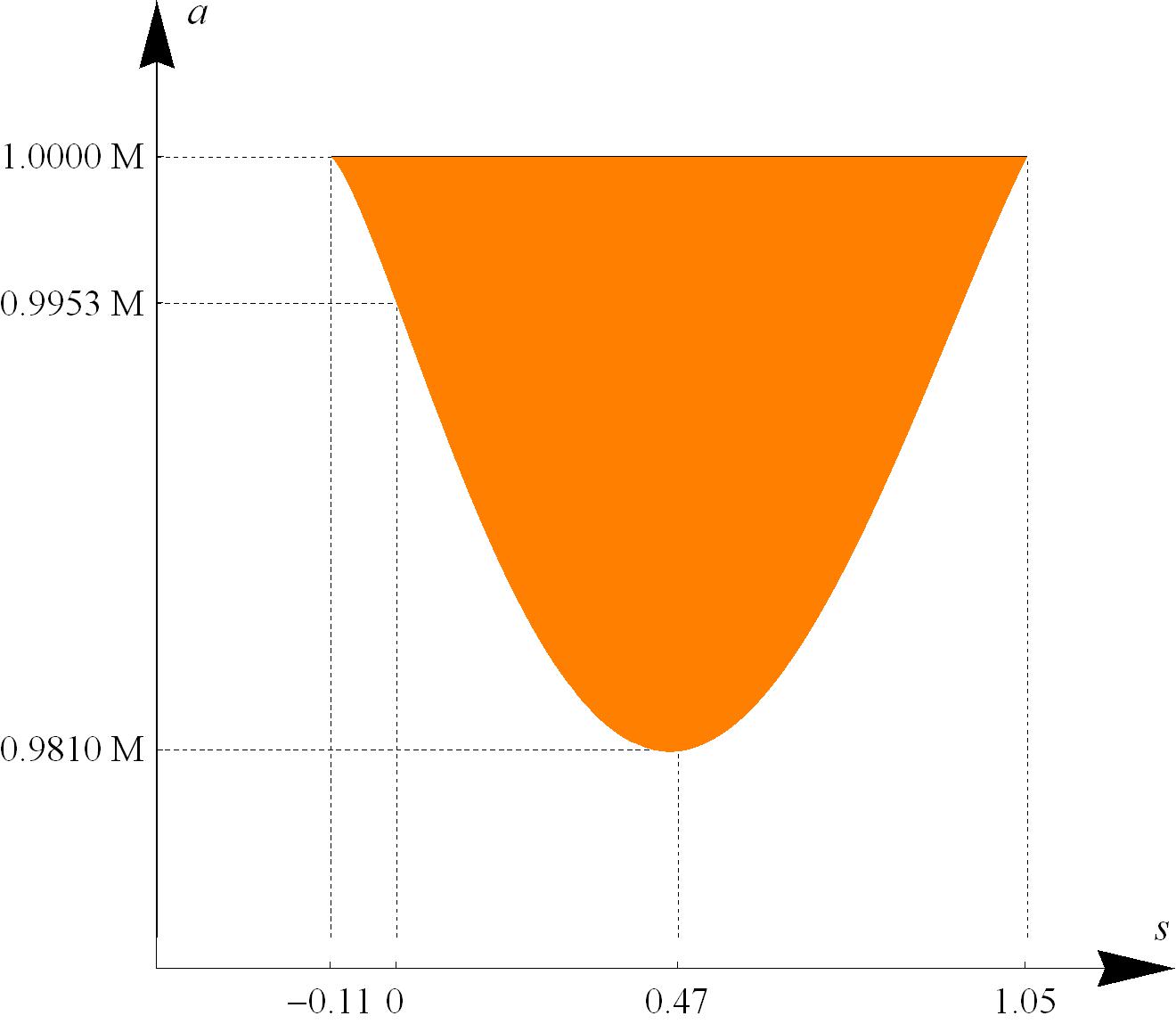}    
\end{center} 

\vspace{-0.5cm}

\caption{\small
Domain in an $s-a-$diagram where the Aschenbach effect occurs
in the domain of outer communication of a Kerr black hole.}  \label{fig:ac}
\end{figure}

Eliminating from the two equations
\begin{equation}
\dfrac{d \mathcal{V}{}_+}{dr} =0 \, ,  \quad
\dfrac{d^2 \mathcal{V}{}_+}{dr^2} =0 
\end{equation}
the radius coordinate $r$ and solving for $a$ gives us the critical value
$a = a_c (s)$ of the black-hole spin where the Aschenbach effect sets in.
This can be done only numerically. Figure \ref{fig:ac} shows the result.
In this figure the region where $\mathcal{V}{}_+$ increases with $r$
is shown shaded (in orange). The lower boundary curve of this region 
gives the critical black-hole spin, $a_c (s)$, as a function
of $s$. For each value of $s$, the non-monotonic behavior 
of $\mathcal{V}{}_+$ as a function of $r$ is present for all values
$a > a_c(s)$; at $a_c(s)$ this function has an inflection point. We have cut 
off the shaded (orange) region at $a =M$ because at the moment
we are only considering the black-hole case, not the naked-singularity case.

\begin{figure}[H]
\begin{center} 
\includegraphics[width=18.5cm]{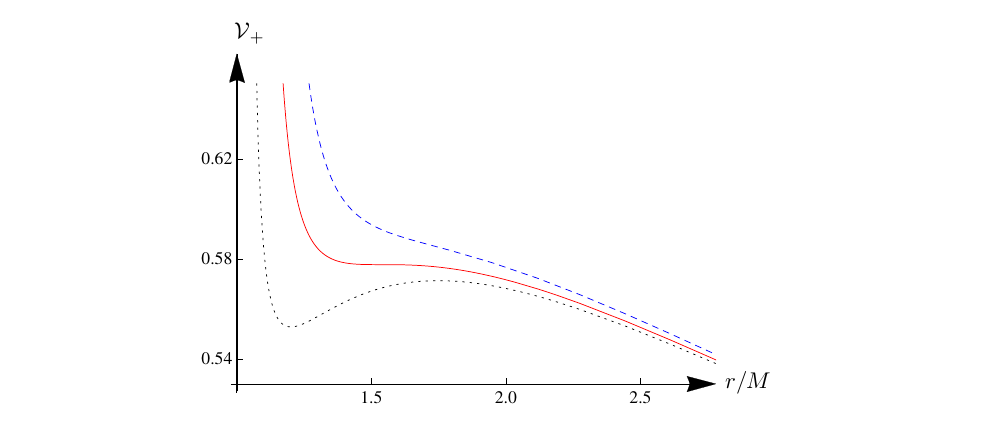}    

\vspace{-0.25cm}

\caption{\small
$\mathcal{V}{}_+$ as a function of $r$ for spinless particles in a Kerr spacetime
with $a= a_c (0) \approx 0.9953 \, M$ (solid, red), $a = 0.9900 \, M$ (dashed, blue)
and $a= 0.9990 \, M$ (dotted, black).} \label{fig:minmax0}
\end{center} 
\end{figure}

We see that for spinless particles ($s=0$) the Aschenbach effect 
sets in at $a_c(0) \approx 0.9953 \, M$, which is the result 
that was found in the original work by 
Aschenbach \cite{Aschenbach2004,Aschenbach2006}. 
This is also illustrated in Figure \ref{fig:minmax0} where 
$\mathcal{V}{}_+$ is shown for a spinless particle in the
equatorial plane of a Kerr spacetime: If $a = a_c(0) \approx 0.9953 M$ 
there is an inflection point; if $a < a_c (0)$ there is no extremum and
if $a > a_c (0)$ there is a minimum-maximum structure. 
If $s$ is negative (i.e., if the spin of the particle is anti-parallel to the spin 
of the black hole), Figure \ref{fig:ac} demonstrates that $a_c (s)$ 
is bigger than $a_c (0)$. If $s$ is positive (i.e., if the spin of the particle
is parallel to the spin of the black hole), $a_c (s)$ can be smaller than 
$a_c (0)$. The minimum value of $a$ where the Aschenbach effect 
could set in is at $a_c (s) \approx 0.9810 \, M$ which happens for a particle
spin of $s \approx 0.47$. So we see that for spinning particles the critical value of $a$ may be reduced only by about one percent in comparison to
the case of spinless particles.  An interesting result which we can also
read from Figure \ref{fig:ac} is that for large (positive or negative) spins
there is no Aschenbach effect around black holes: We have to restrict to
\begin{equation}
-0.11 < s < 1.05 \, ,
\end{equation}
otherwise a non-monotonic behavior of $\mathcal{V}{}_+$ occurs 
only for naked singularities ($a> M$) but not for black holes.

For small values of the spin parameter $s$ we may restrict to a 
Taylor approximation of  $a_c (s)$, 
\begin{equation}
a_c (s) = \Big( 0.9953 - 0.0517 \, s - 0.0164 \, s^2 + O (s^3) \Big) M  
\, ,
\end{equation}
which again was found numerically. The radius coordinate where the 
inflection point occurs is at
\begin{equation}
r_c (s) = \Big( 1.5363 + 1.2155 \, s -  3.9655 \, s^2 + O (s^3) \Big) M  
\, .
\end{equation}

\vspace{-0.5cm}

\begin{figure}[H]
\begin{center} 
\includegraphics[width=16.25cm]{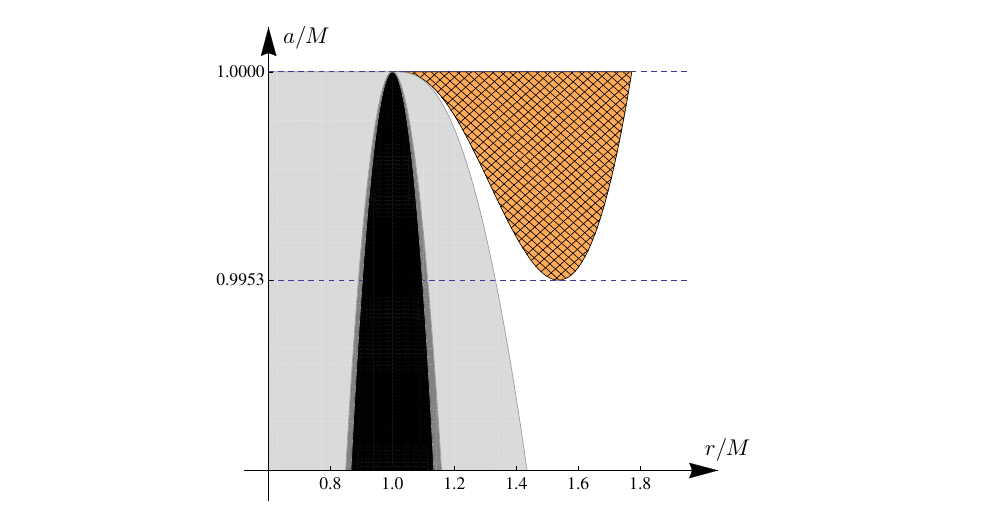}    

\vspace{-0.5cm}

\caption{\small
Region outside of a Kerr black hole where $d \mathcal{V}{}_+/dr >0$ 
for a spinless particle, $s=0$.}
\label{fig:RegBHVp0}
\end{center} 
\end{figure}

\vspace{-0.7cm}

\begin{figure}[H]
\begin{center} 
\includegraphics[width=18.0cm]{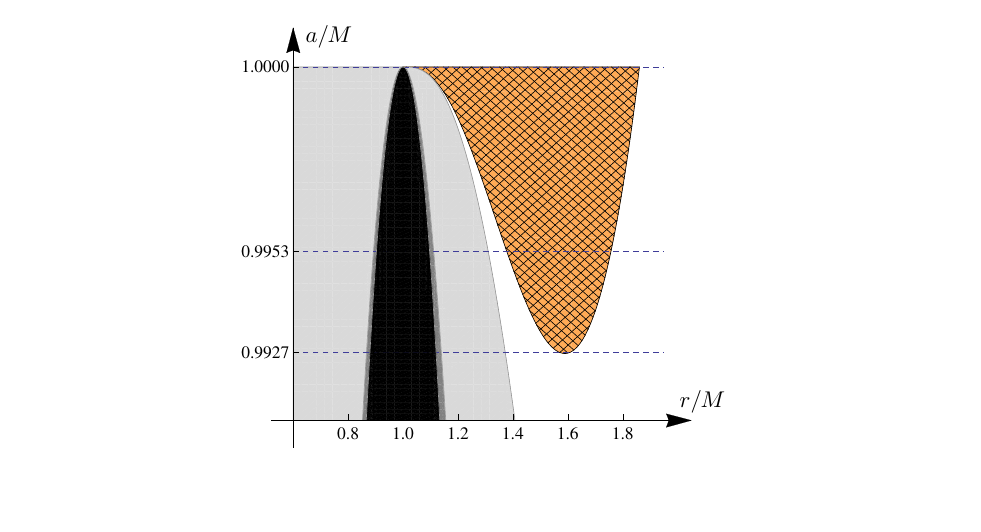}    

\vspace{-1.5cm}

\caption{\small
Region outside of a Kerr black hole where $d \mathcal{V}{}_+/dr >0$ 
for a particle with spin parallel to the spin of the black hole, $s=0.05$.}
\label{fig:RegBHVpp}
\end{center} 
\end{figure}

\begin{figure}[H]
\begin{center} 
\includegraphics[width=18cm]{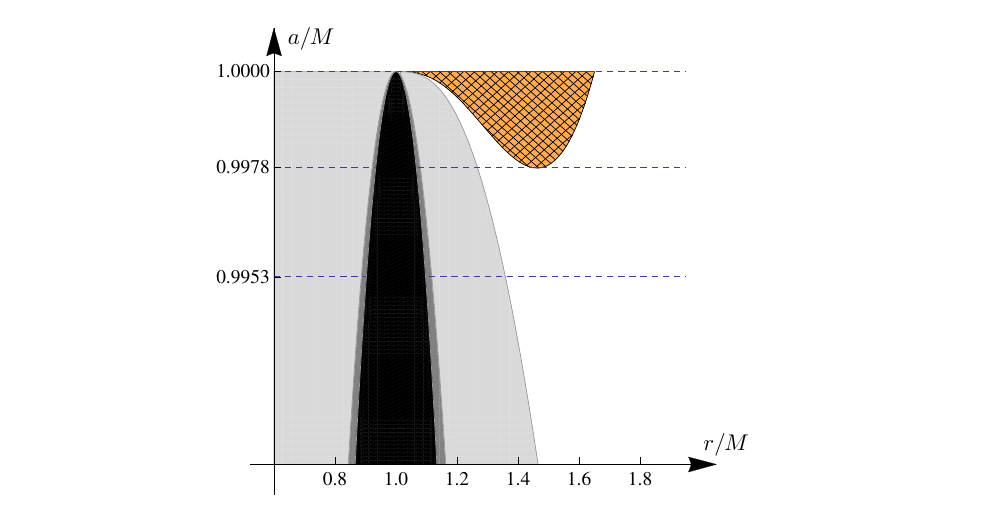}    

\vspace{-0.75cm}

\caption{\small
Region outside of a Kerr black hole where $d \mathcal{V}{}_+/dr >0$ 
for a particle with spin anti-parallel to the spin of the black hole, $s=-0.05$.}
\label{fig:RegBHVpn}
\end{center} 
\end{figure}

\begin{figure}[H]
\begin{center} 
\includegraphics[width=18cm]{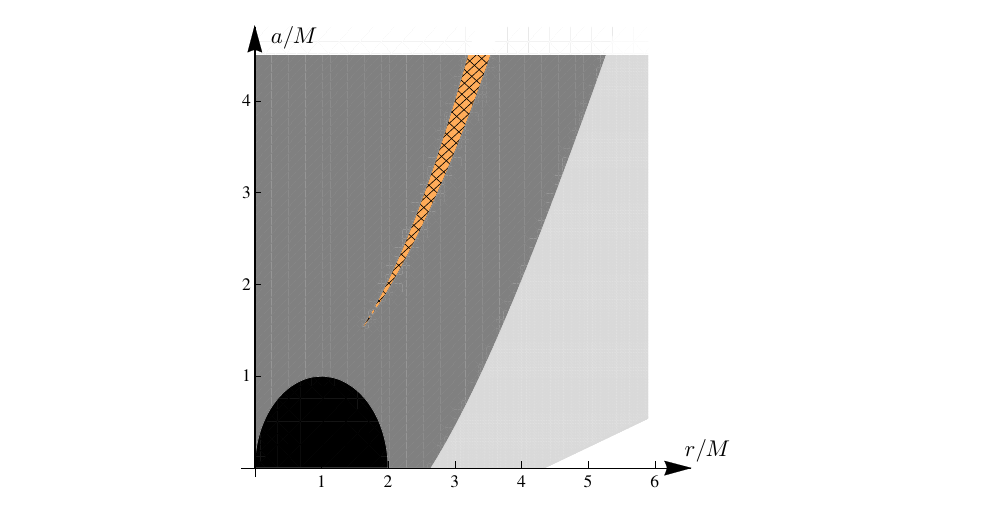}    

\vspace{-0.5cm}

\caption{\small
Entire domain where $| \mathcal{V}{}_- |$ is increasing with $r$, for 
$s=-0.8$}.
\label{fig:RegVnn}
\end{center} 
\end{figure}

It is also instructive to view the domain where $\mathcal{V}{}_+$ is
increasing with $r$ in an $r-a$ diagram. This is shown in Figures
\ref{fig:RegBHVp0}, \ref{fig:RegBHVpp} and \ref{fig:RegBHVpn}
for $s=0$, $s>0$ and $s<0$, respectively. 
In all three pictures the parameter $a$
is restricted to values $a<M$ and only the domain of outer 
communication is considered. We have already said that in 
this domain $\mathcal{V}{}_+$ is always positive. 
In the pictures the region where 
$d \mathcal{V}{}_+/dr > 0$ is shown cross-hatched (in orange).  
For $s>0$ this region may be bigger than for the spinless case,
whereas for $s<0$ it is always smaller. 
In these three figures, and also in the following 
Figs. \ref{fig:RegVp0}, \ref{fig:RegVpp}, \ref{fig:RegVpn} and
\ref{fig:RegVnn}, we have shown the region between the horizons 
in black, the region where circular orbits are unstable in light gray
and the region where circular orbits do not exist at all in
dark gray. We see that in the situations of 
Figs. \ref{fig:RegBHVp0}, \ref{fig:RegBHVpp} 
and \ref{fig:RegBHVpn}
all circular orbits are stable in the domain where the Aschenbach
effect takes place. Also note that all these orbits are inside
the ergoregion whose boundary intersects the equatorial plane at
$r = 2 m$.

\begin{figure}[H]
\begin{center} 
\includegraphics[width=17cm]{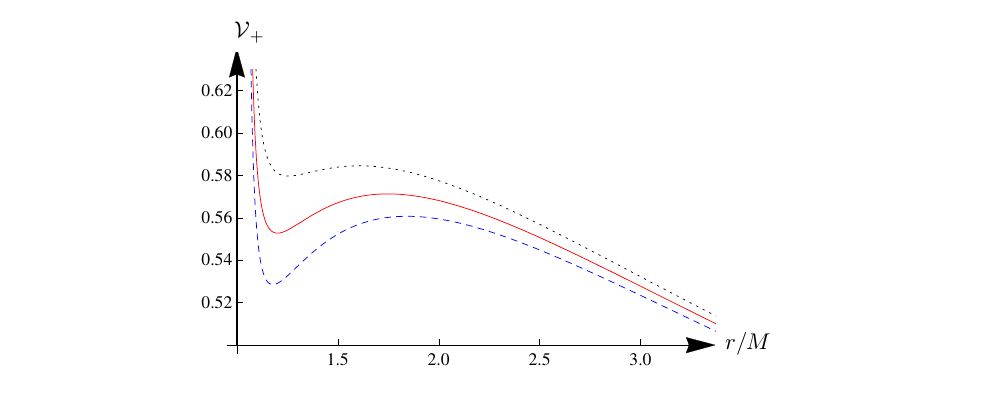}    

\vspace{-0.55cm}

\caption{\small
$\mathcal{V}{}_+$ as a function of $r$, for $a= 0.997 \, M$ 
and $s=0$ (solid, red), $s=0.05$ (dashed, blue)) and $s=-0.05$
(dotted, black).} \label{fig:minmax1}
\end{center} 
\end{figure}

In Figure \ref{fig:minmax1} we show the non-monotonic behavior of
$\mathcal{V}{}_+$ for parallel and anti-parallel particle spin in comparison
to the case of a spinless particle.  

\vspace{-0.5cm}

\vspace{-0.6cm}

\begin{figure}[H]
\begin{center} 
\includegraphics[width=15.5cm]{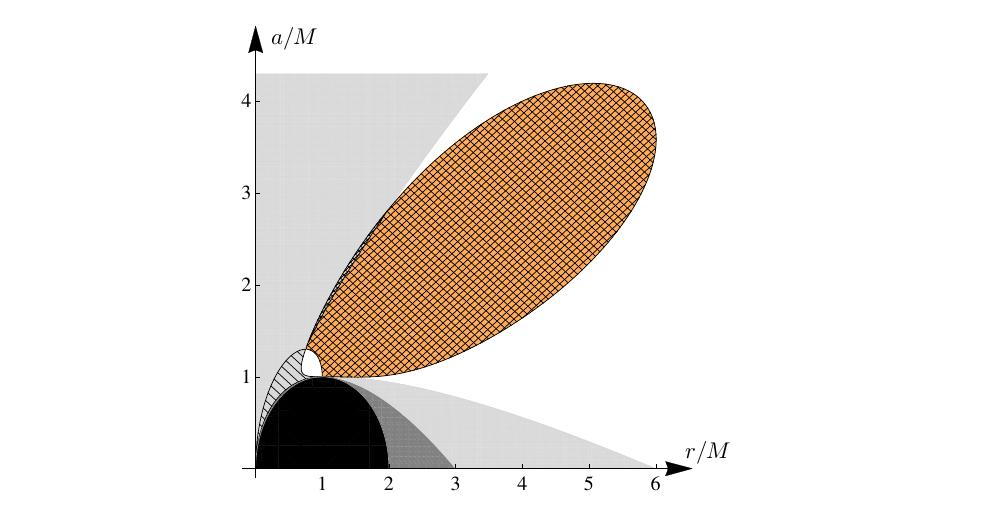}    

\vspace{-0.95cm}

\caption{\small
Entire domain where $| \mathcal{V}{}_+ |$ is increasing with $r$, for 
$s=0$}.
\label{fig:RegVp0}
\end{center} 
\end{figure}

\vspace{-0.93cm}

Having clarified what happens for black holes in the domain of outer
communication, we now briefly discuss the Aschenbach effect in the
entire parameter space, i.e., we allow $a$ to take values bigger than
$M$ and we also consider, in the black-hole case, the domain inside
the inner horizon. (Between the horizons no timelike circular orbits
can exist.) As in the equatorial plane the passage through $r=0$
is blocked by the ring singularity, we do not consider the domain
where $r<0$.

\begin{figure}[H]
\begin{center} 
\includegraphics[width=19cm]{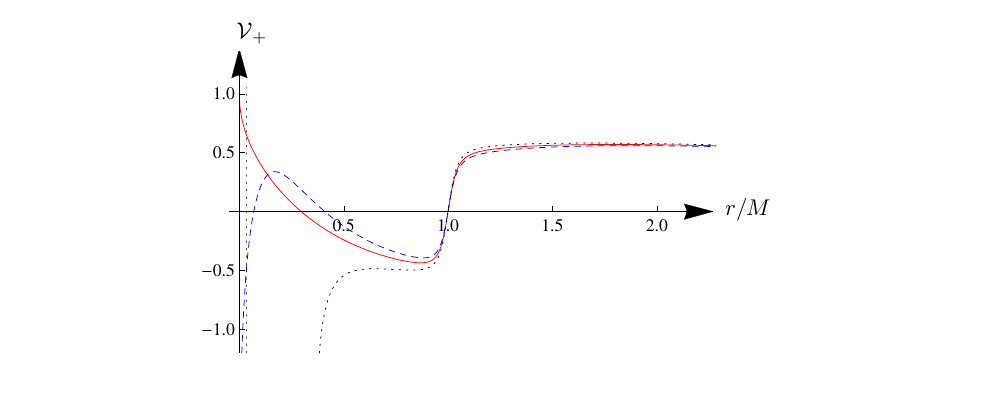}    

\vspace{-0.5cm}

\caption{\small
$\mathcal{V}{}_+$ versus $r$, for $a= 1.001 \, M$ 
and $s=0$ (solid, red), $s=0.05$ (dashed, blue) and $s=-0.05$
(dotted, black).
} 
\label{fig:minmax2}
\end{center} 
\end{figure}

We first consider the $+$ branch of solutions. 
Figs.~\ref{fig:RegVp0}, \ref{fig:RegVpp} and \ref{fig:RegVpn} 
show the entire domain where  the Aschenbach effect takes place
for $s=0$, $s>0$ and $s<0$, respectively. This domain is 
characterized by the properties that the discriminant $D$ is positive,
$\mathcal{V}{}_+$ lies between $-1$ and 1, and 
$| \mathcal{V}{}_+ |$ increases with $r$. We see that this is the
union of two domains: On the first one, shown cross-hatched
(in orange), $\mathcal{V}{}_+$ is positive, i.e., the orbits are
co-rotating. On the second one, shown hatched, $\mathcal{V}{}_+$
is negative, i.e., the orbits are counter-rotating. Only the cross-hatched
(orange) region has an intersection with the domain of outer communication 
of black holes; this intersection was shown, enlarged, in Figures
\ref{fig:RegBHVp0}, \ref{fig:RegBHVpp} and \ref{fig:RegBHVpn}.
We see that the Aschenbach effect is largely taking place in the
naked-singularity domain.  In Figure \ref{fig:minmax2} we show 
the non-monotonic behavior of $\mathcal{V}{}_+$ in a naked-singularity
spacetime for parallel and anti-parallel particle spin in comparison 
to the case of a spinless particle.

\vspace{-0.6cm}

\begin{figure}[H]
\begin{center} 
\includegraphics[width=15.5cm]{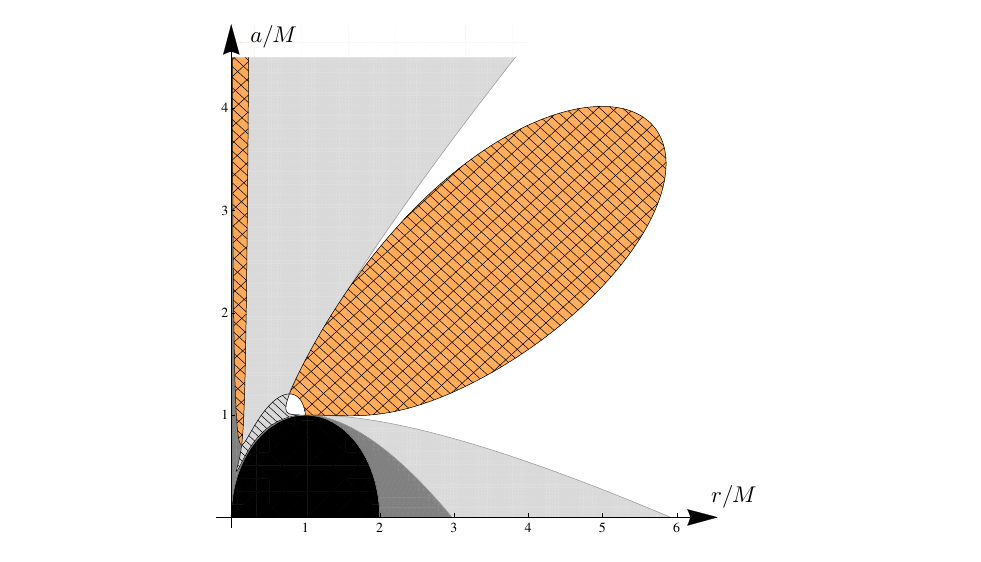}    

\vspace{-0.75cm}

\caption{\small
Entire domain where $| \mathcal{V}{}_+ |$ is increasing with $r$, for 
$s=0.05$}.
\label{fig:RegVpp}
\end{center} 
\end{figure}

\vspace{-0.5cm}

\begin{figure}[H]
\begin{center} 
\includegraphics[width=17cm]{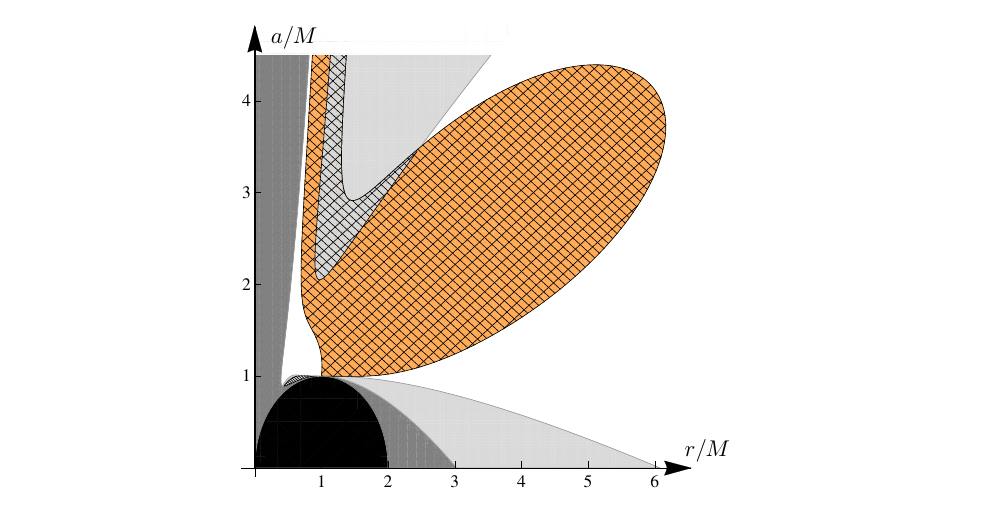}    

\vspace{-0.5cm}

\caption{\small
Entire domain where $| \mathcal{V}{}_+ |$ is increasing with $r$, for 
$s=-0.05$}.
\label{fig:RegVpn}
\end{center} 
\end{figure}

\vspace{-0.4cm}

Whereas in the domain of outer communication of a black hole
the velocity $| \mathcal{V}{}_- |$ is always decreasing with $r$,
this is no longer true if we consider the entire parameter space. 
For any value of $a>1$ and negative spin values in a certain 
interval that depends on $a$, the velocity $| \mathcal{V}{}_- | $
is monotonically increasing on a certain $r$ interval. This interval 
is bounded on the lower side by a radius value where 
$\mathcal{V}{}_-$ is zero which means that the particles are hovering 
at rest with respect to the ZAMOs, and on the upper side by 
a limiting radius where $\mathcal{V}{}_-=-1$ which corresponds
to a counter-rotating orbit at the speed of light. There is no 
minimum-maximum structure. This region is shown
in Figure \ref{fig:RegVnn} cross-hatched (in orange). For the
picture we have chosen the rather big value of $s = - 0.8$ 
because for smaller values the region would be so narrow that
it could hardly be seen.

\vspace{-0.2cm}

\vspace{-0.4cm}

\section{Conclusions}
Up to now, we have much better information on the masses than the spins of 
black-hole candidates. In our view, the astrophysical relevance of the 
Aschenbach effect is in the fact that it provides a method of determining 
the spins of (some) black holes because its occurrence
is associated with a certain parametric resonance of vertical and radial epicyclic 
oscillations. The latter are observable as peaks in the power spectrum
emitted from matter orbiting the black hole.

The main motivation of the present paper is in the fact that we wanted 
to investigate if and how Aschenbach's results are modified if the radiating 
source is spinning. If we think of a hot spot, orbiting the black hole in
an accretion disk, this modification might be non-negligible, in particular if 
we want to rely on the value of $a_c$ up to several digits after the decimal 
point.  The results obtained in this paper could, of course, also be applied 
to a neutron star orbiting a sufficiently massive black hole. To be sure, 
as we worked with the Mathisson-Papapetrou-Dixon equations throughout, 
in any case one has to be aware of the fact that we restricted to situations 
where the test-particle approximation is valid. 

Our analysis was based on the exact (i.e., fully analytical) solutions for the  
orbital velocity in the Locally Non-Rotating Frame (LNRF), for a spinning test 
particle. Thereupon, we have numerically determined the critical value of 
the black-hole spin parameter, $a_c$, where the Aschenbach effect sets in, 
in dependence of the spin parameter $s$ of the test particle. This is only a 
first, but crucially important, step towards our goal. The second step would 
be to investigate the influence of the particle's spin on the parametric 
resonances. We are planning to do this in a follow-up paper.

We have investigated in this paper not only the case of black holes but also 
of naked singularities. The latter are, of course, much more speculative
than black holes. However, we believe that it should be kept in mind that
the Aschenbach effect occurs also for naked singularities, and even in a 
much wider parameter range than for black holes, and that, for the discussion
of parametric resonances, the case of a naked singularity should not be
completely ignored.
  
\vspace{-0.2cm}

\section*{Acknowledgment}
VP is grateful to Old{\v r}ich Semer{\' a}k for helpful discussions
on the motion of spinning particles in general relativity. JK wishes to thank
ZARM, Bremen, for hospitality where part of this work was done. Moreover, VP 
gratefully acknowledges support from Deutsche Forschungsgemeinschaft 
within the Research Training Group 1620 ``Models of Gravity".  

\vspace{-0.2cm}


\begin{thebibliography}{10}
\bibitem{Aschenbach2004}
B.~Aschenbach,
Astron. \& Astrophys. {\bf 425}, 1075 (2004).

\bibitem{Aschenbach2006}
B.~Aschenbach,
Chinese J. Astron. Astrophys. {\bf 6}, S1, 221 (2006).

\bibitem{Shapiro} S. D. Shapiro and S. A. Teukolsky, \emph{Black Holes, 
White Dwarfs and Neutron Stars} (Wiley, New York, 1983).
 
\bibitem{StuchlikSlanyToeroekAbramowicz2005}
Z.~Stuchl{\'i}k, P.~Slan{\'y}, G.~T{\"o}r{\"o}k, and M.~A. Abramowicz,
Phys. Rev. D {\bf 71}, 024037 (2005).

\bibitem{MuellerAschenbach2007}
A.~M{\"u}ller and B.~Aschenbach,
Class.\ Quant.\ Grav. {\bf 24}, 2637 (2007).

\bibitem{SlanyStuchlik2007}
P.~Slan{\' y} and Z.~Stuchl{\'\i}k,
arXiv:0709.0803 [gr-qc] (2007).

\bibitem{zdenek2014}
Z.~Stuchl{\'\i}k, M.~Blaschke and P.~Slan{\' y},
in Z. Stuchl{\'\i}k, G. T{\"o}r{\"o}k and T. Pech{\'a}{\v c}ek (eds.),
Proceedings of RAGtime 2014, Opava, p. 173, (2014)

\bibitem{TursunovStuchlikKolos2016}
A.~Tursunov, Z.~Stuchl{\'\i}k and M.~Kolo{\v s},
Phys. Rev. D {\bf 93}, 084012 (2016).

\bibitem{Mathisson1937}
M.~Mathisson, Acta Phys. Pol. {\bf 6}, 163 (1937).

\bibitem{Papapetrou1951}
A.~Papapetrou, Proc. R. Soc. A {\bf 209}, 248 (1951).

\bibitem{Dixon1964}
W. G. Dixon, Nuovo Cim. {\bf 34}, 317 (1964).

\bibitem{Rasband1973}
S.~N.~Rasband,
Phys. Rev. Lett. {\bf 30}, 111 (1973).

\bibitem{TodFeliceCalvani1976}
K.~P.~Tod, F.~de~Felice and M.~Calvani,
Nuovo Cim. B {\bf 34}, 365 (1976).

\bibitem{Tulczyjew1959} W. Tulczyjew, 
Acta Phys. Pol. {\bf 18}, 393 (1959).

\bibitem{Dixon1970} W. G. Dixon, 
Proc. Roy. Soc. Lond. A {\bf 314}, 499 (1970).

\bibitem{Frenkel1926} J. Frenkel, 
Z. Phys. 37, 243 (1926), Nature {\bf 117}, 653 (1926).

\bibitem{Pirani1956} F. A. E. Pirani, 
Acta Phys. Pol. {\bf 15}, 389 (1956).

\bibitem{Moller1949} C.~M{\o}ller, 
Ann. Inst. Henri Poincar{\'e} {\bf11}, 215 (1949)

\bibitem{BoyerLindquist1967}
R.~Boyer and R.~W. Lindquist, J. Math. Phys. {\bf 8}, 265 (1967). 

\bibitem{zamo}
J.~M.~Bardeen, W.~H.~Press and S.~A.~Teukolsky,
Astrophys.\ J.\  {\bf 178}, 347 (1972).

\bibitem{LNRF}
C.~W.~Misner, K.~S.~Thorne and J.~A~. Wheeler,  \emph{Gravitation}
(Freeman, San Francisco, 1973). 

\bibitem{SaijoEtAl1998}
M.~Saijo, K.-i.~Maeda, M.~Shibata and Y.~Mino,
Phys. Rev. D {\bf 58}, 064005 (1998).

\bibitem{LukesEtAl2017}
G.~Lukes-Gerakopoulos, E.~Harms, S.~Bernuzzi and A.~Nagar,
Phys. Rev. D {\bf 96}, 064051 (2017).

\bibitem{CostaLukesSemerak2018}
L.~F.~O. Costa, G.~Lukes-Gerakopoulos and O.~Semer{\' a}k,
Phys. Rev. D {\bf 97}, 084023 (2018).

\bibitem{ChiconeMashhoonPunsly2005}
C.~Chicone, B.~Mashhoon and B.~Punsly,
Phys. Lett. A {\bf 343}, 1 (2005).

\bibitem{HarmsEtAl2016}
E.~Harms, G.~Lukes-Gerakopoulos, S.~Bernuzzi and A.~Nagar,
Phys. Rev. D {\bf 94}, 104010 (2016).

\end{thebibliography}

\end{document}